\documentclass{article}
\usepackage[dvips]{graphicx}
\makeatletter
\def\@biblabel#1{#1.}
\makeatother

\begin{document}
{\Large\bf
Anisotropic dielectric constant of the parent antiferromagnet 
Bi$_2$Sr$_2M$Cu$_2$O$_8$ ($M$=Dy, Y and Er) single crystals
}

\vspace*{4mm}
 
T. Takayanagi, T. Kitajima, T. Takemura and I. Terasaki

Department of Applied Physics, Waseda University,
Tokyo 169-8555, JAPAN

\vspace*{5mm}

{\bf Abstract:}
The anisotropic dielectric constants of 
the parent antiferromagnet 
Bi$_2$Sr$_2M$Cu$_2$O$_8$ ($M$=Dy, Y and Er) single crystals were 
measured from 80 to 300 K. 
The in-plane dielectric constant is found to be 
very huge (10$^4$-10$^5$).
This suggests a remnant of the Fremi surface of the parent antiferromagnet.
The out-of-plane dielectric constant is 50-200,
which is three orders of magnitude smaller 
than the in-plane one.
A significant anomaly is that a similar 
out-of-plane dielectric constant
is observed in superconducting samples.


\vspace*{7mm}


{\large\bf INTRODUCTION}

In a low hole density, the CuO$_2$ plane shows high resistivity
and antiferromagnetic (AF) order at low temperature, 
which is called a parent AF insulator.
With doping, an insulator-metal transition (IMT)
arises, and the system changes from the AF insulator
to a superconductor.
For IMT, the dielectric constant $\varepsilon$
is of great importance
in the sense that it provides a measure of localization
length in the insulator.
Chen {\it et al.} \cite{Chen} have first pointed out the
importance of $\varepsilon$ in the studies 
of high-$T_c$ cuprates (HTSC).
However, they studied $\varepsilon$ only for La$_2$CuO$_{4+\delta}$,
which has various structural phase transitions that might
affect  $\varepsilon$ seriously.
Another problem is that  they studied $\varepsilon$ only near 4.2 K,
although the resistivity anisotropy was
strongly dependent on temperature.
Thus it should be further examined to study $\varepsilon$
for other HTSC over a wider temperature range.

We have been studying the charge transport of 
the parent insulator Bi$_2$Sr$_2M$Cu$_2$O$_8$ 
($M$=Y and rare-earth) \cite{Kitajima}.
In this proceedings we report on measurements and analyses
of the anisotropic dielectric constants
from 80 to 300 K.

{\large\bf EXPERIMENTAL}

Single crystals of Bi$_2$Sr$_2M$Cu$_2$O$_8$ ($M$=Dy,Y and Er) 
were grown by a self-flux method. 
The growth conditions and the sample characterization 
were described in Ref \cite{Kitajima}.
The resistivity was measured using a four-probe technique,
and a ring configuration was used for the out-of-plane direction.
The dielectric constants were measured
with a two-probe technique using a lock-in amplifier
(Stanford Research SR630 and SR844).
A typical contact resistance was 50-100 $\Omega$
for the in-plane direction,
and 1-10 $\Omega$ for the out-of-plane direction.
Thus the measurement along the in-plane direction
is less accurate near room temperature,
where the contact resistance 
becomes comparable with the sample resistance.
Detailed information on the measurements will be written elsewhere.

All the samples of Bi$_2$Sr$_2M$Cu$_2$O$_8$
were insulating, and
the doping levels of the as-grown crystals
were slightly different for different $M$.
We do not yet understand the $M$ dependence, 
but the melting points
and/or the liquidus lines may depend on $M$
to give a slight variation in composition.
Thus crystals with different $M$'s 
act as a set of parent insulators
with slightly different doping levels.
We estimated the hole concentration per Cu ($p$)
by measuring the room-temperature thermopower \cite{Tallon}.
With good reproducibility,
$M$=Dy was nearly undoped ($p$=0-0.02),
and $M$=Er and Y were slightly doped ($p$=0.02-0.04).
The doping levels (and the measurement results)
were nearly the same between $M$=Er and Y, 
we discuss the data for $M$= Dy and Er below. 

\begin{figure}
\begin{center}
 \includegraphics[width=15cm,clip]{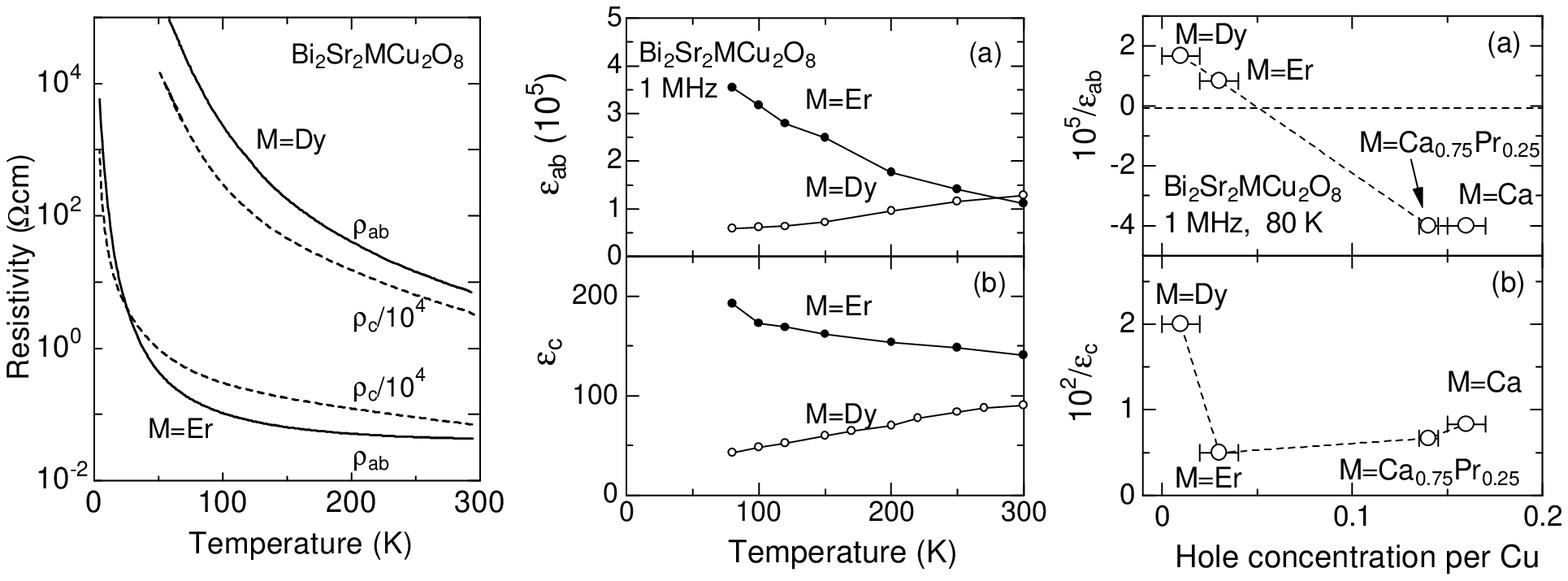}
\end{center}
{\small 
Fig.~1 \quad 
The in-plane resistivity ($\rho_{ab}$) and 
the out-of-plane  resistivity ($\rho_c$) 
of Bi$_2$Sr$_2M$Cu$_2$O$_8$ single crystals.
(left panel)\\
Fig.~2 \quad 
The dielectric constant
of Bi$_2$Sr$_2M$Cu$_2$O$_8$ single crystals.
(middle panel)
(a) The in-plane direction ($\varepsilon_{ab}$)
and (b) the out-of-plane direction ($\varepsilon_{c}$).\\
Fig.~3 \quad
The reciprocal of the  dielectric constant
$1/\varepsilon$ plotted as a function of hole concentration per Cu.
(right panel).
(a) The in-plane direction ($10^5/\varepsilon_{ab}$)
and (b) the out-of-plane direction ($10^2/\varepsilon_{c}$).
For the superconducting samples ($M$=Ca and Ca$_{0.75}$Pr$_{0.25}$), 
see text.
}
\end{figure}

{\large\bf RESULTS AND DISCUSSION}

Figure 1 shows the in-plane resistivity ($\rho_{ab}$) and
the out-of-plane resistivity ($\rho_c$)
for $M$=Er and Dy.
Reflecting the different doping levels,
both $\rho_{ab}$ and $\rho_c$ are  larger
for $M$=Dy than for $M$=Er.
The temperature dependence is also different between 
$M$=Dy and Er.
In particular, $\rho_{ab}$ for $M$=Er is
nearly independent of temperature at 300 K,
which indicates that the in-plane conduction
is nearly metallic.
It should be noted here that $\rho_c/\rho_{ab}$ is
strongly dependent on temperature and the doping levels,
which suggests the confinement behavior
in the AF insulator \cite{Kitajima}.

Figure 2 shows 
the in-plane dielectric constant ($\varepsilon_{ab}$)
and the out-of-plane dielectric constant ($\varepsilon_c$) 
for $M$=Er and Dy at 1 MHz.
Both $\varepsilon_{ab}$ and $\varepsilon_c$
are larger for $M$=Er than $M$=Dy,
which indicates that the sample for
$M$=Er is closer to IMT boundary.
It should be emphasized that $\varepsilon_{ab}$
is as huge as 10$^4$-10$^5$.
We think that the huge $\varepsilon_{ab}$
comes from an electronic origin,
because (1) $\varepsilon_{ab}$ is very sensitive 
to the doping levels and (2) the dielectric loss 
Im $\varepsilon_{ab} \propto 1/\rho_{ab}$
is large compared with conventional ferroelectric materials.
The charge order or the variable range hopping
may be an origin of the huge $\varepsilon_{ab}$.
Thus we may say that the huge $\varepsilon_{ab}$ is a remnant of the 
the Fermi surface calculated by band theories.

An important feature is that $\varepsilon_c$ remains positive and finite
in the superconducting samples.
Kitano {\it et al.} \cite{Kitano} found
that $\varepsilon_c$ of Bi$_2$Sr$_2$CaCu$_2$O$_8$ near $T_c$
was 40-50 at 10 GHz,
whereas Terasaki and Tajima \cite{Terasaki} 
measured that it was 120 at 100 MHz.
These values are of the same order of $\varepsilon_c$
for the parent insulators,
and we may say that the out-of-plane conductance
of HTSC is a ``remnant'' of the parent insulator.
Another feature is that the temperature dependence of $\varepsilon_c$
is different between $M$=Er and Dy.
Recently we have found that $\varepsilon_c$
for all the samples, {\it including superconducting ones}, 
can be understood with the Debye description of  
dielectric relaxation \cite{Kitajima2},
which has been used for the analyses of 
the dielectric response of the charge density wave \cite{Cava}.

The reciprocal of the dielectric constant
($1/\varepsilon_{ab}$ and $1/\varepsilon_c$) at 80 K is plotted 
as a function of hole concentration per Cu in Fig. 3.
For comparison, the data for the superconducting samples are also plotted.
$\varepsilon_{c}$ is employed from Refs. \cite{Terasaki,Kitajima2},
and $\varepsilon_{ab}$ is estimated from 
the Drude model as $\varepsilon_{ab} (\omega\to 0) = -(\omega_p/\gamma)^2$,
where $\omega_p$ and $\gamma$ are the plasma frequency and the 
damping factor respectively.
By putting $\hbar\omega_p$=1.1 eV and $\hbar\gamma$=$k_BT$,
we get $\varepsilon_{ab}=-2.5\times10^4$,
which is in the same order of $\varepsilon_{ab}$ for $M$=Dy and Er.
As shown in Fig. 3(a),  $1/\varepsilon_{ab}$ crosses zero near $p$=0.05,
and goes negative in the metallic side.
This is exactly what we see IMT in doped Si.
On the other hand, although $1/\varepsilon_c$ becomes 
smaller for $M$=Er than for $M$=Dy,
$1/\varepsilon_c$ for the superconducting samples is
positive, and stay at the same order.
Thus $\varepsilon_c$ is unlikely to diverge at IMT,
as Chen {\it et al.} previously found
that $\varepsilon_c$ for La$_2$CuO$_{4+\delta}$ does not
diverge at IMT \cite{Chen}.
We should note that the gross feature of Fig.~3 is not largely 
dependent on frequency and temparature,
although the data for 1MHz at 80 K was rather arbitrarily 
selected.
More detailed analysis is in progress.

Chen {\it et al.} pointed out two possibilities 
for the non-divergent $\varepsilon_c$.
One is that IMT in HTSC occurs 
only along the in-plane direction,
and the other is that the heavy effective mass 
along the out-of-plane direction makes 
the effective Bohr radius of a hole
shorter than the $c$-axis length.
Our data favors the former scenario.
According to the latter scenario,
$\varepsilon_{ab}/\varepsilon_c$ would be
equal to the effective mass ratio,
which disagrees with our observation that
$\varepsilon_c$ for Bi$_2$Sr$_2M$Cu$_2$O$_8$ is larger
than $\varepsilon_c$ for La$_2$CuO$_{4+\delta}$.
Thus the non-divergent $\varepsilon_c$
does not solely comes from the anisotropic effective mass,
but from the anomalous conduction mechanism 
such as ``confinement''.

{\large\bf SUMMARY}

In summary, we prepared single crystals of 
Bi$_2$Sr$_2M$Cu$_2$O$_8$ ($M$=Y and rare-earth) 
and measured the anisotropic dielectric constants
$\varepsilon_{ab}$ and $\varepsilon_c$ 
from 80 to 300 K.
The present study has revealed that $\varepsilon_{ab}$ 
(10$^4$-10$^5$)
is about three orders of magnitude larger 
than $\varepsilon_c$ (10$^2$).
The huge $\varepsilon_{ab}$ is a remnant of the Fermi surface,
where the dc conductivity is suppressed 
by the strong correlation or localization.
We have found that $\varepsilon_c$ remains near $10^2$
across the insulator-metal transition,
which means that the transition occurs only 
along the in-plane direction.
This can be a piece of evidence of the confinement
behavior of the parent insulators.

{\bf Acknowledgments.}
This work was partially supported by The Kawakami 
Memorial Foundation, 
and by Waseda University Grant
for Special Research Projects (99A-556).
The authors would like to thank S. Tajima
for the rf-conductivity measurements in Superconductivity
Research Laboratory,
International Superconductivity Technology Center.
They also appreciate T. Itoh, T. Kawata, K. Takahata
Y. Iguchi and T. Sugaya for collaboration.

\end{document}